\documentstyle[12pt]{article}
\newcommand{\tr}{\mathop{\rm tr}\nolimits}
\newcommand{\im}{\mathop{\rm Im}\nolimits}
\newcommand{\Ker}{\mathop{\rm Ker}\nolimits}
\newcommand{\adx}{\mathop{\rm ad \, X}\nolimits}
\newcommand{\adX}{\adx}
\begin{document}

\begin{titlepage}

\title{Truncation of functional relations in the XXZ model.}
\author{A. A. Belavin, S. Yu. Gubanov, B. L. Feigin  \\ {} \\
              {\it Landau Institute for Theoretical Physics} \\
              {\it Chernogolovka, Institutsky Prospekt 12,}\\
              {\it Moscow region, 142432,} \\
              {\it Russia}}
\date{hep-th/0008011}  
    
\end{titlepage}	      

\maketitle

\begin{abstract}

The integrable XXZ model with a special open boundary condition is considered. 
 We  study Sklyanin transfer matrices after quantum group 
reduction in roots of unity. In this case Sklyanin transfer matrices 
satisfy a closed system of  truncated functional equations.
 The algebraic reason for the truncation is found.The important role 
 in proving of the result is performed by Zamolodchikov algebra introduced 
 in the paper.
\end{abstract}

\section{Introduction.}

We consider the integrable XXZ model with a special open boundary condition.
Its Hamiltonian
$$
H_{XXZ} = \sum^{N-1}_{n=1}[\sigma^{+}_{n}\sigma^{-}_{n+1}
+\sigma^{-}_{n}\sigma^{+}_{n+1}
+\frac{\cosh(\eta)}{2}\sigma^{z}_{n}\sigma^{z}_{n+1}
+\frac{\sinh(\eta)}{2}(\sigma^{z}_{n} - \sigma^{z}_{n+1})].
$$
This Hamiltonian is invariant under  quantum algebra $U_q(sl(2))$ ,
whose generators $X$, $Y$ and $H$ satisfy the following commutation relations
$$
[H,X]=X, [H,Y]=-Y, [X,Y]=[2H]_q,
$$
where the function $[x]_q \equiv \frac{q^x - q^{-x}}{q - q^{-1}}$
 and $q \equiv e^{\eta}$. 
 We have following representation for $X$, $Y$, and $H$ in the terms of
Pauli matrices :
\begin{eqnarray}
\nonumber
X & = & \sum_{n=1}^{N} q^{\frac{1}{2}(\sigma^{z}_{1}+\ldots+\sigma^{z}_{n-1})}
\sigma^{+}_{n}q^{-\frac{1}{2}(\sigma^{z}_{n+1}+\ldots+\sigma^{z}_{N})}, \\
\nonumber
Y & = & \sum_{n=1}^{N} q^{\frac{1}{2}(\sigma^{z}_{1}+\ldots+\sigma^{z}_{n-1})}
\sigma^{-}_{n}q^{-\frac{1}{2}(\sigma^{z}_{n+1}+\ldots+\sigma^{z}_{N})}, \\
\nonumber
H & = & \sum_{n=1}^{N}\frac{\sigma^{z}_{n}}{2}.
\end{eqnarray}
We  concentrate on the case  where $q^{p+1}=-1$. 
It is known \cite {Lus},\cite {PS} that if $q^{p+1}=-1$, then $X^{p+1}=0$ and
$Y^{p+1}=0$; 
  we can therefore consider the subfactor space
$$
V_p = \Ker X / \im X^p.
$$
The result of  the restricting  the XXZ model on $V_p$ is  the
Minimal Model of Integrable Lattice Theory $LM(p,p+1)$ \cite {PS},
~\cite {BelStr}.
The thermodynamic limit is the ordinary Minimal Model of CFT 
 $M(p,p+1)$  ~\cite {BPZ}  with the Virasoro central charge 
 $c=1-\frac{6}{p(p+1)}$ \cite {PS}.

The XXZ model is integrable \cite {ABBBQ}, and as  shown 
by Sklyanin \cite {Skl}, there exists a family of transfer matrices which
commute with the Hamiltonian and between themselves.
The Sklyanin transfer matrices are also $U_q(sl(2))$ 
invariant ~\cite{SklKul}.
Therefore, they can  also be restricted on $V_p$ .As shown in ~\cite{BelStr},
 a result of this restriction is the truncation of the fusion functional
 relations \cite {Zhou} for the transfer matrices.
 This statement was proved in \cite{BelStr} using $T$-$Q$ Baxter equation.
In this work we  prove this fact more directly and
algebraically.
Namely, we  prove that if $q^{p+1}=-1$, then the Sklyanin transfer matrix
$t_{p/2}(u)$ with the spin $j=p/2$ in theauxiliary space vanishes after the
restriction on $V_p$  because
\begin{eqnarray}
t_{p/2}(u) & = & X^p M + N X,
\label{Xp}
\end{eqnarray}
where $ M $ and $N$ are some operators in the quantum space.\\
In another words,
$$
t_{p/2}(u)=0 \pmod {\Ker X / \im X^{p}} .
$$

Then as it follows from Eq. (\ref{Xp}) that the fusion 
functional relations 
are truncated and transformed to a system of functional equations that 
can be used to obtain  the eigenvalues of the transfer matrices.

This paper is organized as follows. 

In Sec. {\bf 2}, we recall some basic formulas that 
are necessary for introducing the Sklyanin transfer matrices
 in Sec. {\bf 3}.

In Sec. {\bf 4}, we show how to obtain the  operator $X$ from a Sklyanin 
monodromy matrix. 

It will allow us then to obtain  commutation relations between the generators
 of the  quantum algebra $U_q(\hat{sl}(2))$ and the Sklyanin monodromy
  matrices in Sec.{\bf 5}.

In Sec. {\bf 6, 7, 8}, we  express the $L$ operators in  terms 
of the generators of the Zamolodchikov algebra and obtain commutation relations 
between  these new operators and the generators of the quantum
 algebra $U_q(\hat{sl}(2))$ .
 
In Sec. {\bf 9}, we rewrite the Sklyanin transfer matrix in  the 
new variables.

In Sec. {\bf 10}, we introduce the operator $\adx$.

In Sec. {\bf 11}, we write down some useful properties 
of $\adX$.

In Sec-s. {\bf 12}and{\bf 13} we prove ~(\ref{Xp}) which is 
our {\bf main statement} \\
({\sl the $\adX$ theorem}).

\section{Yang--Baxter equation.}

The integrable structure of XXZ  is confined in the well-known equation
\begin{eqnarray}
R^{a n}_{b m}(u-v) L^{b}_{c}(u) L^{m}_{k}(v) & = &
   L^{n}_{m}(v) L^{a}_{b}(u) R^{b m}_{c k}(u-v),
\label{YB}
\end{eqnarray}
where $L$ is a  monodromy matrix whose entries 
$ L_n^m $ ($n,m=1,...,2j+1$) are operators acting in the
 so-called  quantum space, which is  a tensor product
 of $ N $  two-dimensional linear spaces in our case.The dimension $2j+1$ of 
the so-called auxiliary space can be different, strictly speaking
 we should write  something like $ (L_j)_n^m $ instead of $ L_n^m $
 and $ (R_{j_1j_2})^{a n}_{b m}$ instead of $ R^{a n}_{b m} $, but
 sometimes for simplicity   we do  not do that and
 assume that everything is understood from the context.
 The $R_{j_1j_2}$-matrix acts in a tensor product of two auxiliary spaces  
 and satisfies to the Yang--Baxter equation 
$$
R^{a n}_{b m}(u-v) R^{b p}_{c r}(u) R^{m r}_{k s}(v)  = 
   R^{n p}_{m r}(v) R^{a r}_{b s}(u) R^{b m}_{c k}(u-v)
$$
and  the unitarity relation 
$$
(R_{j_1j_2})^{a m}_{c k}(u)(R_{j_1j_2})^{c k}_{b n}(-u)  = 
 \phi_{j_1j_2}(u) \phi_{j_1j_2}(-u) \delta^a_b \delta^m_n.
$$
The monodromy matrix $L$ can be  realized through  an N-uple
 product of $R$-matrixes.
$$
L(u) =  R_{N}(u) \ldots R_{1}(u),
$$
here a comultiplier $R_n$ stands for the $R$-matrix acting in a tenzor 
product of the auxiliary space and of the $n$-th comultiplier of the
quantum space.
 We need the related matrix $\bar L^t(u)$, 
$$
\bar L^t(u)  =  R_{1}(u) \ldots R_{N}(u) \; \sim  \; L^{-1}(-u).
$$
It is easy to show that
$$
\bar L_j^t(u)  =   \xi_j^N(-u) L_j^{-1}(-u) 
$$
where $ \xi_j(u) \equiv \phi_{1/2j}(u) \phi_{1/2j}(-u)$.
Using this equality, it is possible to obtain the  formulas
\begin{eqnarray}
\label{dubl}
R^{a n}_{b m}(u-v) \bar L^{m}_{k}(v) \bar L^{b}_{c}(u)  & = & 
  \bar L^{a}_{b}(u) \bar L^{n}_{m}(v) R^{b m}_{c k}(u-v)  \\ 
\nonumber
L^a_c(u) \bar L^m_k(v) R^{c n}_{b k}(u+v) & = & 
R^{a k}_{c m}(u+v) \bar L^k_n(v) L^c_b(u) 
\end{eqnarray}
 In our case, the $R$-matrix 
$$
R(u) =
\left(
\begin{array}{cccc}
\sinh(u+(\frac{1}{2}+\hat{H})\eta) & \sinh(\eta) \hat{F} \\
\sinh(\eta) \hat{E} & \sinh(u+(\frac{1}{2}-\hat{H})\eta) \\
\end{array}
\right).
$$

The $\hat{E}$, $\hat{F}$, $\hat{H}$ are generators 
of  the  quantum algebra $U_q(sl(2))$:
$[\hat{H}, \hat{E}] = \hat{E}$, $[\hat{H}, \hat{F}] = - \hat{F}$,
$[\hat{E}, \hat{F}] = [2 \hat{H}]_q$.\\
In the representation with the spin $j$ we have
$$
\begin{array}{cccc}
\pi_j(\hat{H})_{mn} = (j+1-n) \: \delta_{m,n},  & m,\: n = 1,\: 2, \: \ldots, \: 2j+1. \\
\pi_j(\hat{E})_{mn} = \omega_{m} \: \delta_{m, n-1}, &
\pi_j(\hat{F})_{mn} = \omega_{n} \: \delta_{m-1, n}. \\
\end{array}
$$
where
$$
\omega_{n}  \equiv \sqrt{[n]_q [2j+1-n]_q}.
$$
The spin  $j$ takes values from the  set ($0, \frac12, 1, \ldots)$.

\section{Sklyanin transfer matrix.}
It was shown by Sklyanin in \cite{Skl}  that if transfer matrices 
are defined as
\begin{eqnarray}
\nonumber
t_{j}(u) & = & \tr_{\pi_j}(e^{- 2(u+\eta)\hat{H}} L_j(u) e^{2 \hat{H} u} 
\bar L_j^t(u)) \\
\nonumber
         & \equiv & \sum^{2j+1}_{n,k=1} e^{2(n-k)u}q^{-2(j+1-n)} 
	 (L_j)^n_k(u) \bar (L_j)^n_k(u),
\end{eqnarray}
where, as  above
$$
(\bar L^m_n)^{j_N \ldots j_1}_{i_N \ldots i_1} =
 (L^m_n)^{i_N \ldots i_1}_{j_N \ldots j_1},
$$
then the transfer matrices commute between themselves,
 $$
 [t_{j_1}(u_1), t_{j_2}(u_2)]=0
 $$
  and with the 
Hamiltonian,
 $$
 [t_j(u), H_{XXZ}]=0
 $$
  The basic transfer matrix $t_{1/2}(u)$ 
with the spin of  the auxiliary space $j=1/2$ is related to the  
 $XXZ$ Hamiltonian by
$$
H_{XXZ} = \left.\frac{\sinh(\eta)}{2}\frac{d \log t_{1/2}(u)}{du} \right|_{u=0}
-\frac{\sinh^2(\eta)}{2 \cosh(\eta)}-\frac{N}{2} \cosh(\eta).
$$

\section{Behavior of $L$ as $u \to \pm \infty$.}
We consider the monodromy matrix in the representation with the spin $1/2$.
It is a $2\times 2$ matrix whose entries are some operators.
Using the expression of the monodromy matrix as an N-uple product of $R$-matrices
 with same auxilary and different quantum spaces it is
easy to obtain its behavior as $u \to \pm \infty$:
\begin{eqnarray}
\nonumber
L(u \to + \infty)  & = &
 2^{-N} e^{(u+\eta/2)(N-1)}
\left(
\begin{array}{cccc}
e^{u+\eta/2} q^{H}  &   (q-q^{-1})X_0\\
 (q-q^{-1}) X  & e^{u+\eta/2} q^{-H}  \\
\end{array}
\right) \\
\nonumber
L(u \to  -\infty)  & = &
(-1)^N 2^{-N} e^{-(u+\eta/2)(N-1)}
\left(
\begin{array}{cccc}
e^{-(u+\eta/2)} q^{-H}  &  -(q-q^{-1}) Y\\
-(q-q^{-1}) Y_0  & e^{-(u+\eta/2)} q^{H}  \\
\end{array}
\right).
\end{eqnarray}

The explicit expressions for $X_0$ and $Y_0$ in terms  of Pauli matrices 
can be derived from  those for $X$ and $Y$ by changing $q \to q^{-1}$ 
and $\sigma^{\pm} \to \sigma^{\mp}$.
The $X$, $Y$, $X_0$, $Y_0$ and $H$ are generators of the 
 quantum affine algebra $U_q( \hat{sl}(2))$. Namely,
if we denote:
$x_1 \equiv X$,  $y_0 \equiv Y_0$, $x_0 \equiv X_0$, $y_1 \equiv Y$,
 $h_0 = -h_1  \equiv H$,
then the relations
$[x_i,y_i]=[2H]_q$, $[x_i,y_j]=0$, $[H,x_0]=-x_0$, 
 $[H,y_1]=-y_1$, $[H,x_1]=x_1$ and  $[H,y_0]=y_0$ are fulfilled as well as
the Serre relations
$$
\left( ad_q x_i \right)^3 x_j=0, \quad \left( ad_q y_i \right)^3 y_j=0,
$$
where
$$
\left( ad_q x_i \right)^3 x_j \equiv
\left( x_i \right)^3 x_j
- [3]_q \left( x_i \right)^2 x_j x_i
+ [3]_q x_i x_j \left(x_i \right)^2
- x_j \left( x_i \right)^3.
$$

\section{Transformation of $L^m_n$ and $\bar L^m_n$ 
under the quantum group action.}

We consider the Eq-s.~(\ref{YB}) and~(\ref{dubl})
 in the limit as $u \to \pm \infty$.
We  take the monodromy matrix $L(u)$ in the representation
 with the spin $j=1/2$.We take the second 
 monodromy matrix $L(v)$  in the representation
  with an arbitrary spin $j$.
The $L(u)$ matrix in the limit yields the  generators
 of the  algebra $U_q( \hat{sl}(2))$ with respect to the 
 formulas in the previous section.
We thus obtain the following commutation relations between 
the $L^m_n$, $\bar L^m_n$, and the generators of $U_q(\hat{sl}(2))$.
We here  write only those containing $X$ and $q^{\pm H}$:
\begin{eqnarray}
\label{HL}
q^H  L^n_k &=& q^{n-k} L^n_k q^H,  \\ 
\nonumber
q^H \bar L^n_k &=& q^{k-n} \bar L^n_k q^H, 
\end{eqnarray}

\begin{eqnarray}
\label{XL}
 X L^n_k &-& q^{2(j+1) - k - n} L^n_k X =\\
\nonumber
 \omega_{k-1} \lambda q^{j+1-n} \:  L^n_{k-1} q^{-H}  
&-& \omega_{n} \lambda q^{j+2-k} \:  L_{n+1}^k q^{H},
\end{eqnarray}
  
\begin{eqnarray}
\label{XL1}
X \bar L^n_k &-& q^{k+n-2(j+1)} \bar L^n_k X =\\
 \nonumber
    \omega_{n-1} \lambda^{-1} q^{k-j-1} \bar L_{n-1}^k q^{ H}
   &-&\omega_{k} \lambda^{-1} q^{n-j-2} \bar L^n_{k+1} q^{-H}, 
\end{eqnarray}
where $\lambda \equiv e^v$ .

These commutation relations can be used to prove~(\ref{Xp}).
However, this way is  rather difficult, and we chose another one.
Equations (\ref{HL}) and (\ref{XL}) show that elements of the
monodromy matrix $L$ are transformed under  the quantum group 
 like a composition of  two representations of it.
We therefore introduce a new algebra (which is just the Zamolodchikov 
algebra) such that each element $L^m_n$ or $\bar L^m_n$ is a product
of two generators  and  Eq-s. (\ref{YB}),(\ref{HL}) 
and (\ref{XL}) follow from the  defining relations of this new algebra.
Rewriting the Sklyanin transfer matrix $t_{p/2}(u)$ in terms of the
generators of the new algebra we can easily prove our main statement
(\ref{Xp}).

\section{$L^m_n$ and $\bar L^m_n$ as a composition
 of elements of the Zamolodchikov algebra.}

We introduce the Zamolodchikov algebra generated
by the operators  $\theta_j^n(u)$, $\theta_n^j(u)$, $\bar \theta_j^n(u)$ and 
$\bar \theta_n^j(u)$ ,where $j$ is any positive half integer and
 $n=1,\ldots ,2j+1$,with 
the following defining relations:
\begin{eqnarray}
  (R_{j_1j_2})^{a n}_{b m}(u-v) \theta_{j_1}^b(u) \theta_{j_2}^m(v) &=&
  \phi_{j_1j_2}(u-v)\theta_{j_2}^n(v) \theta_{j_1}^a(u),  
\nonumber \\
(R_{j_1j_2})^{b m}_{a n}(u-v)\theta_m^{j_2}(v) \theta_b^{j_1}(u) &=&
 \phi_{j_1j_2}(u-v) \theta_a^{j_1}(u) \theta_n^{j_2}(v), 
\label{Z1} \\ 
  \theta_a^{j_1}(u) \theta_{j_2}^m(v) &=&
  \tau_{j_1j_2}(u-v) \theta_{j_2}^m(v) \theta_a^{j_1}(u),\nonumber
\end{eqnarray}  

\begin{eqnarray}
  (R_{j_1j_2})^{a n}_{b m}(u-v) \bar \theta_{j_2}^m(v)
   \bar \theta_{j_1}^b(u) & = &
 \phi_{j_1j_2}(u-v) \bar \theta_{j_1}^a(u) \bar \theta_{j_2}^n(v),
\nonumber \\
(R _{j_1j_2})^{b m}_{a n}(u-v) 
\bar \theta_b^{j_1}(u) \bar \theta_m^{j_2}(v) & = &
  \phi_{j_1j_2}(u-v) \bar \theta_n^{j_2}(v) \bar \theta_a^{j_1}(u), 
\label{Z2}   
\\ \nonumber
  \bar \theta_{j_2}^m(v) \bar \theta_a^{j_1}(u) & = &
  \tau_{j_1j_2}(u-v) \bar \theta_a^{j_1}(u) \bar \theta_{j2}^m(v),
\end{eqnarray}  

\begin{eqnarray}
\nonumber
  (R_{j_1j_2})^{a n}_{b m}(u+v) \bar \theta_{j_2}^n(v) \theta_{j_1}^b(u)
 & = & \chi_{j_1j_2}(u+v)
 \theta_{j_1}^a(u) \bar \theta_{j_2}^m(v), \\
\nonumber
(R_{j_1j_2})^{b m}_{a n}(u+v) \theta_b^{j1}(u) \bar \theta_n^{j2}(v) & = & 
  \bar\chi_{j_1j_2}(u+v) \bar\theta_m^{j_2}(v)\theta_a^{j_1}(v),
 \\ \theta_{j1}^a(u) \bar \theta_n^{j_2}(v) & = &
  \rho_{j_1j_2}(u+v) \bar \theta_n^{j_2}(v) \theta_{j_1}^a(u), 
  \label{Z3}  
\\ \nonumber
\theta_n^{j_2}(v) \bar \theta_{j_1}^a(u) & = &
\bar\rho_{j_1j_2}(u+v) \bar \theta_{j1}^a(u) \theta_n^{j2}(v), 
\end{eqnarray}  
where $\phi_{j_1j_2}$, $\tau_{j_1j_2}$, $\rho_{j_1j_2}$ and $\chi_{j_1j_2}$
 are some functions. 
These functions depend on the specific form of $R_{j_1j_2}$. For example
 $\phi_{j_1j_2}$ is defined from the unitarity equation and is equal to the 
 multiplier in the r.h.s. of this equation.We don't know how to obtain the
  form of the three other fuctions  in general case, but
 we can  find  the needed combination of them from the
 self-consistency requirement.
The Zamolodchikov algebra is associative for arbitrary  
 $\phi_{j_1j_2}$, $\tau_{j_1j_2}$, $\rho_{j_1j_2}$  and $\chi_{j_1j_2}$
 provided only that the $R$-matrix satisfies  the triangle equation.
  A similar algebra was considered in~\cite{Miwa}.

It is easy  to verify using (\ref{Z1})--(\ref{Z3}) that  if 
\begin{eqnarray}
\label{LTeta}
(L_j)^a_b(u) & =  &   \theta_j^a(u)\theta_b^j(u), \\
\nonumber
 (\bar L_j)^a_b(u) & = &\bar \theta_b^j(u) \bar \theta_j^a(u),
\end{eqnarray}
then  $L$ and $\bar L$  satisfy the Yang--Baxter equations
(\ref{YB}) and (\ref{dubl}).

\section{Transformation of $\theta_n$, $\theta^n$, $\bar \theta_n$ and 
$\bar \theta^n$ under the quantum group action.}

The following commutation relations
between  $L_a^b$ , $\bar L_a^b$ and the generators of the Zamolodchikov 
algebra,which follow from (\ref{Z1})--(\ref{LTeta}),
are sufficient for our purpose:

\begin{eqnarray}
R^{a n}_{b m}(u-v) L^b_c(u) \theta_j^m(v)  & = &
  \theta_j^n(v) L^a_c(u) \phi(u-v), 
\end{eqnarray}

\begin{eqnarray}
 R^{b m}_{a n}(u-v)\theta_m^j(v) L^c_b(u)& = & 
  \phi(u-v) L^c_a(u) \theta_n^j(v) , 
\end{eqnarray}

\begin{eqnarray}
  R^{a n}_{b m}(u-v) \bar \theta_j^m(v) \bar L^b_c(u) & = &
  \bar L^a_c(u) \bar \theta_j^n(v) \phi(u-v), 
 \end{eqnarray}
 
  \begin{eqnarray}
 R^{b m}_{a n}(u-v) \bar L^c_b(u) \bar  \theta_m^j(v)& = & 
  \phi(u-v)  \bar \theta_n^j(v) \bar L^c_a(u). 
\end{eqnarray}
Here we write $L_{\frac{1}{2}}$ as just  $L$ , $R_{\frac{1}{2} j}$ as $R$ and
 set$ \phi(u)=\phi_{\frac{1}{2} j}(u) \tau_{\frac{1}{2} j}(u) $.
We already know how  to obtain the operator $X$  from  $L(u)$. 
Having taken the limit $u \to \pm \infty$ we can obtain the commutation
 relation between $X$ and $\theta$.
To do this we must know  what the function $\phi(u)$ is equal to.
But the function  $\phi(u)$ depends on the function 
 $\tau_{\frac{1}{2} j}(u), $which is unknown.
Actually, we only need to know the behavior of  $\phi(u)$  in the limit 
$u \to + \infty$ .Taking this limit  we obtain 

\begin{eqnarray}
\nonumber
q^H \theta_j^n(v) & = & \varepsilon(v) \; \;  q^{n-j-1} \theta_j^n(v) q^H, \\
\nonumber
q^{-H} \theta_j^n(v) & = & \varepsilon(v) \; \; q^{j+1-n} 
\theta_j^n(v) q^{-H}, \\
\nonumber
q^H \theta_n^j(v) & = &  \varepsilon(v)^{-1} q^{j+1-n} 
\theta_n^j(v) q^H, \\
\nonumber
q^{-H} \theta_n^j(v) & = & \varepsilon(v)^{-1}  q^{n-j-1}
 \theta_n^j(v) q^{-H},
\end{eqnarray}
where 
$$
\varepsilon (v) \equiv \lim_{u \to + \infty} 
(2 \phi(u-v) e^{-(u-v+\eta/2)}).
$$
 It is obvious that these equations are self-consistent  only if
$$
\varepsilon (v) = 1.
$$
Supposing that this is true we  also obtain the commutation 
relations between $X$ and $\theta$,thus,

\begin{eqnarray}
\nonumber
q^H \theta_j^n & = & q^{n-j-1} \theta_j^n q^H, \\
q^H \theta_n^j & = & q^{j+1-n} \theta_n^j q^H ,\\
\nonumber
q^H \bar \theta_j^n & = & q^{j+1-n} \bar \theta_j^n q^H, \\
\nonumber
q^H \bar \theta_n^j & = & q^{n-j-1} \bar \theta_n^j q^H ,
\end{eqnarray}

\begin{eqnarray}
\nonumber
X\theta_j^n- q^{j+1-n}\theta_j^n X & = & 
- \omega_n \; \lambda \; q \; \theta_j^{n+1} q^H ,\\
X\theta_n^j- q^{j+1-n}\theta_n^j X & = &  
\omega_{n-1} \; \lambda \; \theta_{n-1}^j q^{-H}, \\
\nonumber
X \bar \theta_j^n- q^{n-j-1} \bar \theta_j^n X & = & 
 \omega_{n-1} \lambda^{-1} \bar\theta_j^{n-1} q^H, \\
\nonumber
X \bar \theta_n^j- q^{n-j-1}\bar \theta_n^j X & = & 
-\omega_{n} \lambda^{-1} q^{-1} \bar \theta_{n+1}^j q^{-H}. 
\end{eqnarray}

It is easy to verify that  these formulas are consistent 
with  (\ref{HL})and (\ref{XL})
if (\ref{LTeta}) is taken into account.

\section{New variables $\psi$.}
We can make our formulas more convenient if we introduce the new
variables $ \psi $ by
 
\begin{eqnarray}
\nonumber
\theta_j^n &=& \psi_j^n q^{-n H}, \\
\label{}
\bar \theta_j^n &=& \bar  \psi_j^n q^{n H}, \\
\nonumber
\theta_n^j &=& \psi_n^j q^{-n H}, \\
\nonumber
\bar \theta_n^j &=& \bar  \psi_n^j q^{n H}, 
\end{eqnarray}

\begin{eqnarray}
\nonumber
X \psi_j^n -q^{j+1-2n}\psi_j^nX & = & -\omega_n \lambda q \psi_j^{n+1}, \\
\label{Xpsi}
X \bar \psi_j^n -q^{2n-j-1}\bar \psi_j^nX & = & 
\omega_{n-1} \lambda^{-1} \bar \psi_j^{n-1}, \\
\nonumber
X \psi_n^j -q^{j+1-2n}\psi_n^jX & = & 
\omega_{n-1} \lambda  \psi_{n-1}^j, \\ 
\nonumber
X \bar \psi_n^j -q^{2n-j-1}\bar \psi_n^jX & = & 
-\omega_{n}\lambda^{-1}q^{-1} \bar \psi_{n+1}^j. 
\end{eqnarray}

\section{Transfer matrix in the new variables.}
 We rewrite the Sklyanin transfer matrices 
in the new variables $\psi$.
The result suggests  that the transfer matrices consist 
of two other objects. 
Replacing the old variables with the new ones ,

\begin{eqnarray}
\nonumber
(L_j)^n_k (\bar L_j)^n_k & = & \theta_j^n \theta_k^j
 \bar \theta_k^j \bar \theta_j^n \\
\nonumber
& = & q^{k(j+1-k)}q^{-n(j+1-n)} \psi_j^n \psi_k^j
 \bar \psi_k^j \bar \psi_j^n,
\end{eqnarray}
we obtain 

\begin{eqnarray}
\label{t19}
t_j(u) = \sum^{2j+1}_{n=1}\lambda^{2n}q^{-(n+1)(j+1-n)}\psi_j^n
\left(\sum^{2j+1}_{k=1}\lambda^{-2k}q^{k(j+1-k)}
\psi_k^j \bar \psi_k^j \right)\bar \psi_j^n
\end{eqnarray}
(here and after  $\lambda = e^u$).
We see that the transfer matrix consists of two independent structures.
If we set

\begin{eqnarray}
\label{G}
g^{-}_j(u) & \equiv & \sum^{2j+1}_{k=1}\lambda^{-2k}q^{k(j+1-k)}
\psi_k^j \bar \psi_k^j, \\
\nonumber
g^{+}_j(u) & \equiv & \sum^{2j+1}_{n=1}\lambda^{2n}q^{-(n+1)(j+1-n)}
\psi_j^n\bar \psi_j^n,
\end{eqnarray}
then the transfer matrix has the  form
$$
t_j(u) =  \sum^{2j+1}_{n=1}\lambda^{2n}q^{-(n+1)(j+1-n)}
\psi_j^n \; g^{-}_j(u) \;\bar \psi_j^n. 
$$
The new objects $g^{-}_j(u)$ and $g^{+}_j(u)$ are remarkable
 because they are separately invariant with respect to
 the  quantum algebra $U_q(sl(2))$,
$$
  [g^{\pm}_j(u), U_q(sl(2))] = 0.
$$
 We  prove this formula after introducing of the $\adx$ operator.

\section{ The$\adx$ operator.}

It is convenient to introduce a linear operator( we call it the $\adx$operator)
 such  that the following properties  hold
 
\begin{eqnarray}
\label{P1}
\adx (\Psi +  \Psi') & = & \adx (\Psi) +  \adx (\Psi') 
\end{eqnarray}
and also if $ \Psi $ and $ \Psi' $ have definite degrees,

\begin{eqnarray}
\nonumber
\adx (\Psi \Psi') & = &  \adx (\Psi) \Psi' + q^{\deg(\Psi)}
 \Psi \adx(\Psi'), \\
\label{P2}
\adx (\Psi) & = & X \Psi - q^{\deg(\Psi)} \Psi X, \\
\nonumber
\deg(\Psi \Psi') & = & \deg(\Psi) +  \deg(\Psi'). 
\end{eqnarray}
 By definition we set 
$$
\begin{array}{cccccc}
\adx (\psi_j^n) & =& X \psi_j^n -q^{j+1-2n}\psi_j^nX & = &
 -\omega_n \lambda q \psi_j^{n+1} \\
\adx (\bar \psi_j^n) & = & X \bar \psi_j^n -q^{2n-j-1}
\bar \psi_j^nX & = & \omega_{n-1}
 \lambda^{-1} \bar \psi_j^{n-1} \\
\adx (\psi_n^j) & = & X \psi_n^j -q^{j+1-2n}\psi_n^jX & = & 
\omega_{n-1} \lambda  \psi_{n-1}^j \\ 
\adx (\bar \psi_n^j) & =& X \bar \psi_n^j -q^{2n-j-1}
\bar \psi_n^jX & = & 
-\omega_{n} \lambda^{-1}q^{-1} \bar \psi_{n+1}^j \\
\end{array}
$$
(compare with ~(\ref{Xpsi})).
We define  the  degrees of these operators by
\begin{eqnarray}
\nonumber
\deg(\psi_j^n) & = & j+1-2 n ,\\
\label{}
\deg(\bar \psi_j^n) & = & 2 n-j-1,\\
\nonumber
\deg(\psi_n^j) & = & j+1-2 n ,\\
\nonumber
\deg(\bar \psi_n^j) & = & 2 n-j-1,
\end{eqnarray}
and also

\begin{eqnarray}
\nonumber
\deg(X^N \psi_j^n) & = &\deg(\psi_j^n) -2 N, \\
\label{}
\deg(X^N \bar \psi_j^n) & = & \deg(\bar \psi_j^n) -2 N,\\
\nonumber
\deg(X^N \psi_n^j) & = & \deg(\psi_n^j)+2N, \\
\nonumber
\deg(X^N \bar \psi_n^j) & = &\deg(\bar \psi_n^j)+2N.
\end{eqnarray}

 Pay your attention that applying the operator $X$ to $\psi$ and
 $\bar \psi$ with upper index $n$  decrease their degrees and conversely
 applying to the ones with lower index $n$  increase their degrees.
 It is now easy  to show that $\adx(g^{\pm}_j(u) ) = 0$.

Indeed, 
\begin{eqnarray}
\adx g^{-}_j(u)
 \nonumber
 & = & \sum^{2j+1}_{k=1}\lambda^{-2k}q^{k(j+1-k)}
\left(  \adx (\psi_k^j) \bar \psi_k^j  + 
 q^{j+1-2 k} \psi_k^j \adx( \bar \psi_k^j)  \right) \\
\nonumber
& = &
\sum^{2j+1}_{k=1}\lambda^{-2k}q^{k(j+1-k)}
\left( \omega_{k-1} \lambda \psi_{k-1}^j \bar \psi_k^j  
-  q^{j+1-2 k}  \omega_{k} \lambda^{-1} q^{-1}
  \psi_k^j \bar \psi_{k+1}^j  \right) \\
\nonumber
& = & 0.
\end{eqnarray} 
Because $\deg(g^{\pm}_j(u))=0$, we have  
$X g^{\pm}_j(u)=g^{\pm}_j(u) X$.
In fact, even more general statement 
$$
  [g^{\pm}_j(u), U_q(sl(2))] = 0
$$
is true.

\section{Properties of the $\adx$ operator.}
Using  properties ($\ref{P1}$) and ($\ref{P2}$) of the operator $\adx$
 it is easy to prove  the  identities
\begin{eqnarray}
\nonumber
(\adx)^N(\psi^A \bar \psi^B) & = & \sum^{N}_{n=0}
 q^{n(\deg(\psi^A)+n-N)}C^N_n
(\adx)^{N-n}(\psi^A)(\adx)^{n}(\bar \psi^B), \\
\nonumber
(\adx)^N(\psi_A \bar \psi_B) & = & \sum^{N}_{n=0} q^{n(\deg(\psi^A)-n+N)}C^N_n
(\adx)^{N-n}(\psi_A)(\adx)^{n}(\bar \psi_B), \\
\label{prop}
(\adx)^N(\psi^A) & =  & \sum^N_{n=0} (-1)^n 
q^{n(\deg (\psi^A)+1-N)} C^N_n X^{N-n} \psi^A X^n ,\\
\nonumber
(\adx)^N(\psi_A) & =  & \sum^N_{n=0} (-1)^n 
q^{n(\deg (\psi_A)-1+N)} C^N_n X^{N-n} \psi_A X^n,
\end{eqnarray}
where
$$
C^N_n = \frac{[N]_q !}{[n]_q ! [N-n]_q !}
$$
is the $q$-binomial coefficient.
We can observe a light difference in these formulas for $\psi$ and
$\bar \psi$ with the upper and lower indexes.

\section{Proof of the $\adx$ theorem in terms of $\theta$ operators.}

Using the definition of the operator $\adx$, we can obtain

\begin{eqnarray}
 \psi_j^n = & a_n (\adx)^{n-1}(\psi_j^1), \\
\nonumber
 \psi_n^j = &b_n (\adx)^{2j+1-n}(\psi_{2j+1}^j), 
\end{eqnarray}

\begin{eqnarray}
 \bar \psi_j^n = & c_n (\adx)^{2j+1-n}( \bar \psi_j^{2j+1}),\\
\nonumber
  \bar \psi_n^j= &d_n (\adx)^{n-1}(\bar \psi_1^j), 
\end{eqnarray}
where
$$
\begin{array}{cccccc}

a_n = &\prod^{n-1}_{k=1}(-\lambda^{-1}q^{-1}\omega^{-1}_{n}),& &
b_n = &\prod^{2j+1-n}_{k=1}(\lambda^{-1}\omega^{-1}_{2j+1-k}), \\
c_n = &\prod^{2j+1-n}_{k=1}(\lambda\omega^{-1}_{2j+1-k}),& &
d_n = &\prod^{n-1}_{k=1}(-\; \lambda \; q \; \omega^{-1}_{n}).\\
\end{array}
$$
It is easy to verify that
$$
\begin{array}{cccccc}
a_n c_n =& (-1)^{n-1}\omega^{-1} q^{1-n} \lambda^{2(j+1-n)}, \\
b_n d_n =& (-1)^{n-1}\omega^{-1} q^{n-1} \lambda^{2(n-j-1)},\\
\end{array}
$$
where
$$
\omega = \prod^{2j}_{k=1} \omega_k.
$$
Therefore,

\begin{eqnarray}
\nonumber
 \psi_n^j \bar \psi_n^j & = & 
  (-1)^{n-1}\omega^{-1} q^{n-1} \lambda^{2(n-j-1)}
  (\adx)^{2j+1-n}(\psi_{2j+1}^j)(\adx)^{n-1}(\bar \psi_1^j), \\
\nonumber
  \psi_j^n \bar \psi_j^n & = & (-1)^{n-1}\omega^{-1} q^{1-n} 
  \lambda^{2(j+1-n)}
  (\adx)^{n-1}(\psi_j^{1})(\adx)^{2j+1-n}(\bar \psi_j^{2j+1}).
\end{eqnarray}

Inserting this in  the expression  for $g^{-}_j(u)$, we obtain

\begin{eqnarray}
\nonumber
g^{-}_j(u) & = & \sum^{2j+1}_{k=1}\lambda^{-2k}q^{k(j+1-k)}
\psi_k^j \bar \psi_k^j \\
\nonumber
           & = & \omega^{-1} \lambda^{-2(j+1)}q^j 
\sum^{2j}_{n=0}(-1)^n q^{n(j-n)} 
(\adx)^{2j-n}(\psi_{2j+1}^j)(\adx)^n (\bar \psi_1^j).
\end{eqnarray}

It makes sense to compare this with the  similar formula
$$
(\adx)^{2j}(\psi_{2j+1}^j \bar \psi_1^j) = 
\sum^{2j}_{n=0}q^{-(2j+1)n} q^{n(j-n)} C^{2j}_n 
(\adx)^{2j-n}(\psi_{2j+1}^j)(\adx)^n (\bar \psi_1^j).
$$
Taking into account that if $q^{p+1}=-1$ then $C^{p}_n=1$ 
and $q^{\pm (p+1)n}=(-1)^n$,
 we obtain  the conclusion 
\begin{eqnarray}
\label{g-}
g^{-}_{p/2}(u)=\omega^{-1}\lambda^{-2(p/2+1)}q^{p/2} 
(\adx)^{p}(\psi_{p+1}^{p/2} \bar \psi_1^{p/2}).
\end{eqnarray}

Similarly, we have
\begin{eqnarray}
\label{g+}
g^{+}_j(u) & = & \sum^{2j+1}_{n=1}\lambda^{2n}q^{-(n+1)(j+1-n)}
\psi_j^n\bar \psi_j^n \\
\nonumber
           & = & (-1)^{2j} \omega^{-1} \lambda^{2(j+1)}q^{j(2j+1)}
\sum^{2j}_{n=0}(-1)^{-n} q^{n(j+n)} (\adx)^{2j-n}
(\psi_j^1)(\adx)^n (\bar \psi_j^{2j+1})
\end{eqnarray}
 for $g^{+}_j(u)$;  as above if $q^{p+1}=-1$, we obtain

$$
g^{+}_{p/2}(u)=(-1)^{p}\omega^{-1}\lambda^{2(p/2+1)}q^{p(p+1)/2} 
(\adx)^{p}(\psi_{p/2}^1 \bar \psi_{p/2}^{p+1}).
$$
If $q^{p+1}=-1$,we can  use (\ref{g+}) and $\adx(g^{-}_j(u) ) = 0$ to rewrite 
formula (\ref{t19})
  for the transfer matrix  $t_{p/2}(u)$ if  as
\begin{eqnarray}
\label{MT1}
t_{p/2}(u) & = & (\adx)^{p}(G_{p/2}(u)),
\end{eqnarray}
where we introduce the notation
\begin{eqnarray}
\label{G}
G_{p/2}(u)\equiv  (-1)^{p}\omega^{-1}\lambda^{2(p/2+1)}q^{p(p+1)/2}
\left(\psi_{p/2}^1 \; g^{-}_{p/2}(u) \; \bar \psi_{p/2}^{p+1}\right) 
\end{eqnarray}

Because of  property (\ref{prop}) of $adx$ and because
 $q^{p+1}=-1$ implies $C^{p}_n=1$, we can also rewrite  (\ref{MT1}) as
 
\begin{eqnarray}
\label{MT2}
t_{p/2}(u) & = &\sum^{p}_{n=0} X^{p-n} G_{p/2}(u) X^n.
\end{eqnarray}

 We  have thus proved in fact our {\bf main result}( {\sl the adX theorem}).
In the next section we return  from  the "virtual" variables
$\psi$ and $\bar \psi$ to  the variables $L$ and $\bar L$.

\section{Return to $L$ and $\bar L$.}

 By returning from the variables $\psi$ and $\bar \psi$ 
to the variables $L$ and $\bar L$ in (\ref{G}),
we can easily show that
$$
G_{p/2}(u) = (-1)^{p}\omega^{-1} q^{p^2/2} \sum^{p+1}_{k=1}
\left( \lambda^{2(p/2+1-k)}(L_{p/2})^1_k (\bar L_{p/2})^{p+1}_k \right) q^{-p H}
$$

 Although we have proved the main theorem using  intermediate 
calculations  with the "virtual"  operators $\theta$, which somehow
do not exist, this proof is nevertheless  valid .
Indeed, to prove Eq.(\ref{MT2}), we could permute the 
 operator $X$ with the entire $L$ 
and $\bar L $ to the right   without splitting $L$ and 
 $\bar L$ to  the operators $\theta$ .
Thus acting,  we would not meet these operators at all,
but the result of this permutation must be the same( and Eq.
(\ref{MT2}) must be fulfilled) because of the associativity
 of the Zamolodchikov algebra.
 
 This completes the proof of the $\adx$ theorem.

\section{Acknowledgments.}

We are grateful to M.~Lashkevich and A.~Odesskii
for usefull discussions and also to W.~ Everett fo
the editorial assistance.This work is supported in part
by RBRF-00-15-96579,98-02-16687,99-01-01169,INTAS--97-1312.


\begin{thebibliography}{99}

\bibitem{Lus}
Lusztig E. {\it Contemp. Math.} {\bf 82}  59 (1989)

\bibitem{PS}
 Pasquier V. and   Saleur H.
 { Common structures between finite systems and 
conformal field theories through quantum groups }
{\it Nucl. Phys. B }{\bf 330} 523 (1990)

\bibitem{BelStr}
A. Belavin, Yu. Stroganov;
{  Minimal Models of Integrable Lattice Theory
and Truncated Functional Equations, hep-th/9908050 },
 {\it Phys. Lett. B} {\bf 446} 281 (1999)
 
\bibitem{BPZ}
A.~Belavin, A.~Polyakov, and A.~Zamolodchikov,
{ Infinite conformal symmetry in  two-dimensional quantum field theory}
{\it Nucl.\ Phys.} {\bf~B241} , 333 (1984)

 
 \bibitem{ABBBQ}
  Alcaraz F C,   Barber M N,  Batchelor M T,  Baxter R J and  
 Quispel G R W  Surface exponents of the quantum XXZ, Ashkin--Teller 
and Potts models {\it J. Phys. A: Math. Gen. } {\bf 20}
 6397-6409 (1987)

\bibitem{Skl}
 Sklyanin E K  { Boundary conditions for integrable quantum systems}, 
{\it J. Phys. A: Math. Gen.} {\bf 21}  2375-2389 (1988)

\bibitem{SklKul}
 Kulish P P, Sklyanin E K  { The general $U_q[sl(2)]$ invariant
 $XXZ$ integrable quanrum spin chain}, 
{\it J. Phys. A: Math. Gen.} {\bf 24}  L435-L439 (1991)

\bibitem{Zhou}
Yu-kui Zhou,
{  Fusion hierarchy and finite-size corrections of $U_q(sl(2))$ invariant vertex 
models with open boundaries},
{\it hep-th/9502053} (1995)

\bibitem{Miwa}
Omar Foda, Kenji Iohara, Michio Jimbo, Rinat Kedem, Tetsuji Miwa and Hong Yan,
{Notes on highest weight modules of the elliptic algebra $A_{q,p}(\hat{sl}_2)$},
{\it hep-th/9405058}(1994)

\end{thebibliography}
\end{document}